\documentclass[american,aps,pra,reprint, superscriptaddress, onecolumn]{revtex4-2}
\usepackage{amsmath,amscd,amsthm,amssymb}
\usepackage{mathrsfs}
\usepackage{graphicx}
\usepackage{epsf,epstopdf}
\usepackage[all]{xy}
\usepackage[unicode=true,pdfusetitle, bookmarks=true,bookmarksnumbered=false,bookmarksopen=false, breaklinks=false,pdfborder={0 0 0},backref=false,colorlinks=false] {hyperref}
\hypersetup{colorlinks=true,linkcolor=myurlcolor,citecolor=myurlcolor,urlcolor=myurlcolor}
\usepackage{graphics,epstopdf,graphicx, amsthm, amsmath, amssymb, times, braket, color, bm}
\usepackage{subcaption}

\definecolor{myurlcolor}{rgb}{0,0,0.7}

\theoremstyle{plain}

\providecommand{\theoremname}{Theorem}
\newcommand*{\myproofname}{Proof}

\makeatother

\theoremstyle{definition}

\theoremstyle{remark}

\usepackage{amscd}

\newcommand{\beq}{\begin{equation}}
\newcommand{\eeq}{\end{equation}}
\newcommand{\ba}{\begin{array}}
\newcommand{\ea}{\end{array}}
\newcommand{\bea}{\begin{eqnarray}}
\newcommand{\eea}{\end{eqnarray}}

\begin{document}

\title{Matrix encoding method in variational algorithm of calculating eigenvalues and generalized eigenvalues}

\author{Alexander I. Zenchuk}
\email{zenchuk@itp.ac.ru}
\affiliation{Federal Research Center of Problems of Chemical Physics and Medicinal Chemistry RAS,
Chernogolovka, Moscow reg., 142432, Russia}
\author{Junde Wu}
\email{wjd@zju.edu.cn}
\affiliation{School of Mathematical Sciences, Zhejiang University, Hangzhou 310027, PR~China}

\begin{abstract}
We propose a variational method for constructing the eigenvalues and generalized eigenvalues for an arbitrary $N\times N$ complex matrix. The quantum part of our algorithm is based on  encoding the matrix  elements into the pure state  of a quantum system and  expressing  the loss function with optimization parameters  in terms of  certain probability amplitudes  in the superposition state.  The principal step of this algorithm  is the measurement of the ancilla state 
 that   removes all extra terms from the above superposition and allows to   probabilistically  construct  the required  loss function along with its derivatives with respect to the optimization parameters. These output data are used to find the new values   of optimization parameters for the next iteration of the  loss function   in the gradient optimization method.  
The depth and size  of the circuit for this algorithm are, respectively,   $O(N^2 \log N)$ and $O(\log N)$.
\end{abstract}

\maketitle

\newpage

\section{Introduction}

During last decades, quantum informatics experiences huge progress been involved in such global problems as communication 
\cite{Bose,Werner_1989,BBCJPW,Popescu,MHSWK,Retal,QLNHWZHCLLZDY, PBGWK,PBGWK2,DLMRKBPVZBW},
 cryptography \cite{ZQMLQ,B,TLXQL} and computation \cite{HHL,BWPRWL,SV}.  The later, although been most attractive application of quantum physics, faces set of series obstacles for development due to requirement of organizing  the quantum platform with large number of qubits admitting reliable pairwise interaction between any two of them.  Nevertheless, the theoretical background of quantum computation is rather  well developed \cite{NCh,Preskill}, and there is significant progress in quantum architecture  \cite{Schleich2016,Acin2018}. Intensive research in this direction is stimulated by predicted  exponential speedup of quantum computation algorithms which provides  doubtless advantage  of these algorithms over their classical counterparts. 

Overlooking achievements in development of  quantum algorithms, first of all we refer to Deutsch algorithm \cite{Deutsch} exploring  quantum parallelism, Grover search algorithm \cite{Grover},   Shor's factorization algorithm \cite{Shor1,Shor2}. 
Extensive applicability is deserved by quantum Fourier transform \cite{Coppersmith,WPFLC,NCh} and  phase estimation algorithm \cite{NCh,WWLN}. 

However, at present time, not any classical algorithm has its quantum counterpart. In particular, the quantum optimization algorithms, such as   GRadient Ascent Pulse Engineering (GRAPE) approach   \cite{Khaneja2005}  and   Chopped Random Basis (CRAB) optimization method \cite{Doria2011}, whose application  to  control problems in various physical areas is wildly acknowledged \cite{PP_2023,GlaserJPB2011,Fouquieres2011,PechenTannor2011,Lucarelli2018,Koch2022,Muller2022},   have not been constructed  yet because of the principal obstacles for realizing the computational cycles in quantum algorithms as well as the real-time re-encoding the  optimization parameters of unitary rotations at each iteration during algorithm evaluation. This is the reason that gives rise to intensive development of so-called hybrid algorithms that combines both classical and quantum subroutines. 
 
 Among them, we select the family of variational quantum algorithms which include the classical optimization algorithm  reaching   the optimal  values of the optimization parameters using the loss (or objective) function calculated by means of  quantum algorithm.  Thus, the variational quantum eigensolver allowing  to estimate the ground state energy of Hamiltonians was  proposed in Ref.\cite{PCSYZLAO}.   The variational algorithms for finding eigenvalues and generalized eigenvalues  by scanning the energy interval  were proposed, respectively, in Refs. \cite{XXZ} and   \cite{LSLF}.
The algorithm for constructing the minimal generalized eigenvalue for Hermitian matrices was developed in Ref. \cite{ SWRKWESY}.
In \cite{HJKP},  all generalized eigenvalues of Hermitian operators where  found  successively using the Euclidean time evolution.  The significant promotion was realized  in \cite{ZZW},  where the algorithm for calculating eigenvalues by reducing the matrix to the triangular form was developed for  an arbitrary matrix.
 The gradient optimization method    with effective quantum calculation of  loss function together with its derivatives with respect to optimization parameters  was implemented in variational quantum algorithm for contracting the singular value decomposition \cite{WSW}  (this algorithm was verified  by  its simulating on  the PaddlePaddle Deep Learning Platform \cite{Paddle1, Paddle2, MWW}) 
 and  the generalized eigenvalues \cite{LFYYFZDM}.   We note that   the nonunitarity obstacle for  considering  arbitrary matrices in the above quoted variational algorithms was  solved by 
using  the linear combination of unitaries (LCU) \cite{ZZW,WSW},  by transforming the matrix to the Hermitian form in the case of evolutionary approach \cite{XXZ}, or, alternatively, by  the matrix encoding using specially  reordered matrix elements  \cite{JKC}. The  interest in development of algorithms for solving   generalized eigenvalue problem is supported by its wide application not only in physics, statistic and mathematics, but also in 
new research areas such as  machine learning \cite{DFO,AAF}, quantum chemistry \cite{FH} and algo rithms   operating with large computational data \cite{KLR,TAF}. 

In our paper, we propose the variation algorithm for calculating generalized eigenvalues for arbitrary complex matrices $A$ and $B$ using  transformation of these matrices to the triangular form \cite{LFYYFZDM}. 
The novelty is   encoding these  matrices  into the quantum state of certain quantum system \cite{ZQKW_2024,ZQW_2025}, which, in particular,  allows  to avoid  the problem of  non-unitarity of considered matrices.  Thus, the matrices $A$ and $B$ are considered as a quantum state of a particular quantum system rather than  in terms  of unitary operators as in the most of  traditional variational approaches to the eigenvalue problem cited above.  
We also  apply the similar approach of  matrix  triangularization and encoding to the variational algorithm for calculating the usual eigenvalues of $A$  and concentrate on  some specific features of this algorithm.  
Remark that the method of matrix encoding was first applied to the matrix manipulation algorithms (addition, multiplication, inversion, linear system solver) in \cite{ZQKW_2024,ZBQKW_2024}. The principal novelty and advantage of this method is that it  uses encoding the  matrix elements into the probability amplitudes of the superposition state of certain quantum system and does not require  embedding  the input matrices into the Hermitian operators  with subsequent approximate exponentiation  via Suzuki-Trotter \cite{Trotter,Suzuki,T,AT} or Baker-Champbell-Hausdorff \cite{BC} methods, although the theoretical background for  applying the  latter  methods in matrix algebra is rather well developed \cite{ZhaoL1,HHL,BWPRWL,HHL3,MRTC,NJ}.
Later, the matrix-encoding method was  applied to   the variational algorithm for constructing singular value decomposition of an arbitrary complex matrix \cite{ZQW_2025}. 

In the present work, we combine the quantum algorithm  for calculating the loss function (together with its derivatives with respect to the optimization parameters) at  fixed  values of optimization  parameters with classical gradient algorithm of calculating the values of the optimization parameters for the next iteration step.  Finally, we study the relation among the accuracies for calculating the loss function, eigenvalues and optimization parameters for both algorithms calculating eigenvalues and generalized eigenvalues. All  theoretical arguments and statement are justified by  numerical simulations of  both variational algorithms for $4\times 4$ randomly generated complex matrices.
 
 The paper is organized as follows. The variational  algorithm for calculating  generalized eigenvalues (GEV)  based on matrix encoding into the pure quantum state is proposed  in Sec. \ref{Section:SVD}.  Features of the algorithm constructing  usual eigenvalues  (EV) are discussed in Sec.\ref{Section:ev}.  Characteristics of both  algorithms  including circuit parameters and parameters of the optimization algorithm  together with accuracy of calculation are   considered in Sec.\ref{Section:var}.  Numerical constructing of GEV and EV for  $4\times 4$ matrices and analysis of  accuracy of the calculated eigenvalues  and accuracy of  encoding the optimization parameters  are presented in Sec.\ref{Section:num}. Obtained result are discussed  in Sec.\ref{Section:conclusions}.

\section{Variational  algorithm for constructing generalized eigenvalues}
\label{Section:SVD}
\subsection{Preliminaries} We consider the variational algorithm for constructing GEV for two arbitrary non-degenerated complex  $N\times N$ matrices  $A$ and $B$  assuming $N=2^n$,
\begin{eqnarray}\label{SVD}
A|\psi\rangle=\lambda B  |\psi\rangle,
\end{eqnarray}
where $\lambda$ is the eigenvalue corresponding to the eigenvector $|\psi\rangle$.  
According to the generalized Schur decomposition theory \cite{GL},
 for any matrices $A,B\in \mathbb C^{N\times N}$, there exist the unitary matrices $Q$ and $Z$ such that
\begin{eqnarray}
Q^\dagger AZ=T,\;\; Q^\dagger B Z=S,
\end{eqnarray}
where $T=\{t_{ij}\}, S=\{s_{ij}\} \in \mathbb C^{N\times N}$ are upper triangular matrices.  If, for some $k$, $t_{kk}=s_{kk}=0$, then the appropriate $\lambda_k$  can  be any complex number, otherwise
\begin{eqnarray}\label{lambdak}
\lambda_k = \frac{t_{kk}}{s_{kk}},\;\; s_{kk}\neq 0,\;\;k=0,\dots,N-1.
\end{eqnarray}
We define the loss function as the sum of squares of absolute values of all vanishing elements in $T$ and $S$ \cite{LFYYFZDM}:
\begin{eqnarray}\label{obj}
&&
L(\alpha,\beta) = \sum_{j=1}^{N-1}  \sum_{j=0}^{i-1}  \left( |\langle i| U^T (\alpha) A U(\beta)|j\rangle|^2  +
|\langle i| U^T (\alpha) B U(\beta)|j\rangle|^2\right) =\\\nonumber
&&
 \sum_{j=1}^{N-1}  \sum_{j=0}^{i-1}  \left( |\langle i| T(\alpha,\beta)|j\rangle|^2  +
|\langle i| S (\alpha,\beta)|j\rangle|^2\right)
, 
\end{eqnarray}
where the superscript  $T$  denotes matrix transposition,  $\alpha=\{\alpha_0,\dots,\alpha_{nM}\}$ and $\beta= \{\beta_0,\dots,\beta_{nM}\}$ represent two sets of real optimization parameters,  $M$ is some integer associated with the subroutine used for preparing the unitary transformation $U$ in Sec.\ref{Section:W2W2}.  In Eq. (\ref{obj}), $U$ is an arbitrary $N\times N$ unitary transformation of general form, therefore $U(\alpha)$ and $U(\beta)$ can be regarded as  two arbitrary independent transformations  if only $\alpha$ and $\beta$ are two independent sets of parameters. 
We appeal to  the gradient method of  finding such parameters $\alpha_0$ and $\beta_0$ that minimize  the loss function,
i.e.,
\begin{eqnarray}\label{min}
\min_{\alpha,\beta} L(\alpha,\beta)  = L(\alpha_0,\beta_0).
\end{eqnarray}
It is shown in \cite{LFYYFZDM} that the loss function (\ref{obj}) reaches its global minimum of zero if and only if  $T(\alpha,\beta)$ and $S(\alpha,\beta)$ are upper triangular matrices. 
Then
\begin{eqnarray}\label{UDV}
&&
T=U^T (\alpha_0) A U(\beta_0),\;\;\;S=U^T (\alpha_0) B U(\beta_0),
\\\nonumber
&&
Q^\dagger= U^T(\alpha_0),\;\; Z= U(\beta_0),
\end{eqnarray}
and the generalized eigenvalues can be found using Eqs.(\ref{lambdak}).

\subsection{Quantum algorithm  calculating loss function}
\label{Section:general}

To construct the variational quantum algorithm for calculating generalized eigenvalues, we introduce six $n$-qubit subsystems:  the subsystems $R$ and $C$ serve to enumerate, respectively, the rows and column of $A$ and $B$, the subsystems $\chi$, $\tilde \chi$ and $\psi$, $\tilde\psi$ are needed   to  operate with   $\langle j|U^T(\alpha)$ and $U(\beta)|j\rangle$ in (\ref{obj}).
In addition, the  one-qubit  subsystem $L$ serves to mark the elements of matrices  $A$ and $B$ in the superposition state (see Eq.(8)), and the one-qubit  subsystem $K$ serves as a controlling qubit in  succeeding  controlled operations.  At the last steps of the algorithm we will introduce two  one-qubit ancilae $\tilde B$ and $B$  to properly organize garbage removal via measurement.

First of all, we have to prepare the above mentioned matrices $A=\{a_{ij}: i,j=0,\dots,N-1\}$ and $B=\{b_{ij}: i,j=0,\dots,N-1\}$ for encoding into the state of a quantum system.
To this end, we normalize  these matrices    assuming that  $|a_{00}|^2 + |b_{00}|^2\neq 0$, i.e., replace $A$ and $B$ with
\begin{eqnarray}\label{A}
\tilde A= \frac{A}{\sqrt{ \sum_{i=0}^{N-1}  \sum_{j=0}^{N-1}  ( |a_{ij}|^2+ |b_{ij}|^2)}} , \;\; \tilde B= \frac{B}{\sqrt{ \sum_{i=0}^{N-1}  \sum_{j=0}^{N-1}  ( |a_{ij}|^2+ |b_{ij}|^2)}} .
\end{eqnarray}
Hereafter we do not write tilde over $A$ and $B$.
Now we can  encode the elements  of the matrices $A$ and $B$ into the superposition state of $R$, $C$  and $L$ using, for instance, the state-encoding algorithm developed in Ref.\cite{ZQW_2026} and obtain the following superposition state:
\begin{eqnarray}\label{M}
&&
|\Psi\rangle_{RCL} = |A\rangle  + |B\rangle= 
\sum_{i=0}^{N-1}\sum_{j=0}^{N-1}( a_{ij} |0\rangle_L + b_{ij} |1\rangle_L) |i\rangle_R |j\rangle_C  ,   \;\;  \\\nonumber&&
 \sum_{i=0}^{N-1}\sum_{j=0}^{N-1} (|a_{ij}|^2+|b_{ij}|^2)=1,
\end{eqnarray}
where the normalization is provided by  Eqs.(\ref{A}). In other words, we have quantum access to the matrices  $A$ and $B$ \cite{BLZ}.
Subsystems $\chi$, $\tilde\chi$,   $\psi$,  $\tilde\psi$ and $K$  are  in the ground states initially, i.e., the initial state of the whole system { reads}:
\begin{eqnarray}\label{Phi0}
|\Phi_0\rangle=|\Psi\rangle_{RCL} |0\rangle_\chi |0\rangle_{\tilde\chi}   |0\rangle_{\psi}   |0\rangle_{\tilde\psi}
  |0\rangle_K .
\end{eqnarray}
Hereafter in this paper, the subscript  at  the operator  means the subsystem to which this operator is applied.

\begin{figure}[ht]
\centerline{    \includegraphics[width=0.8\textwidth]{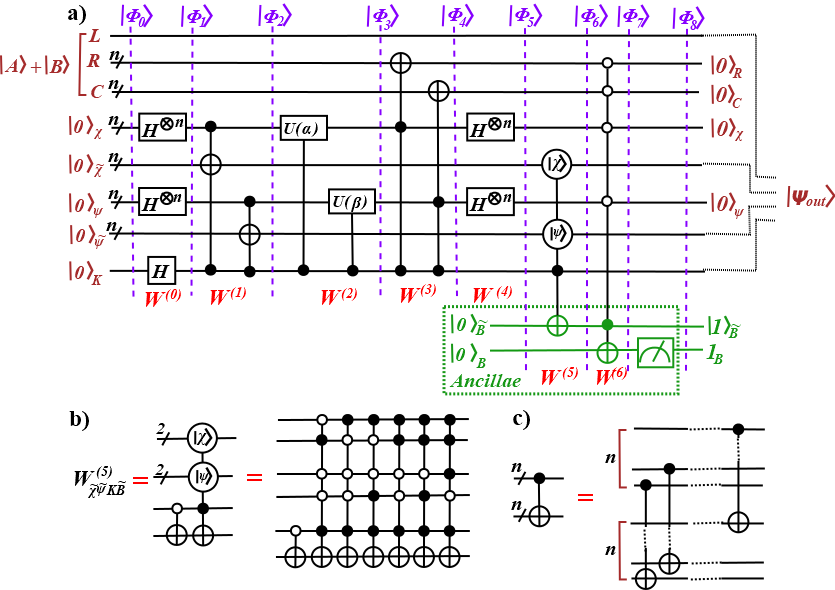}}
    \caption{The quantum circuit  for calculating the loss function.  
  Depending on $M$,  the depth of this circuit can be estimated by either    $O(M\log N)$ or $O(N^2\log N)$ which is defined by the operator  either  $W^{(2)}_{\chi\psi K}$ or $W^{(5)}_{\tilde \chi \tilde \psi K \tilde B}$. (a) The circuit for creating the state $|\Psi_{out}\rangle$ given in (\ref{Phi4}); the operators $W^{(j)}$, $j=0,\dots, 6$ are presented without subscripts  for brevity. (b) The circuit for the operator $W^{(5)}_{\tilde\chi \tilde\psi K\tilde B}$ for the case of $4\times 4$ matrices,  $n=2$. (c) The notation for the multiqubit C-not.}
\label{Fig:SVD}
\end{figure}

The quantum algorithm is  illustrated by  the circuit in Fig.\ref{Fig:SVD}a.
As the first  step, we apply the Hadamard transformation to each  qubit of $\chi$, $\psi$ and $K$ (we denote this set of transformations by $
W^{(0)}_{\chi \psi K}= H_\chi H_\psi  H_K$):
\begin{eqnarray}\label{Phi1}
|\Phi_1\rangle = W^{(0)}_{\chi\psi K} |\Phi_0\rangle =\frac{1}{2^{(2n+1)/2}}
\sum_{k=0}^{N-1}\sum_{l=0}^{N-1} |\Psi\rangle_{RCL} |k\rangle_\chi  |0\rangle_{\tilde \chi}  |l\rangle_\psi  |0\rangle_{\tilde\psi}   ( |0\rangle_K+|1\rangle_K),
\end{eqnarray}
  creating   two  systems of orthonormal states $ |k\rangle_\chi$ and $ |l\rangle_\psi$, 
 $k,l=0,\dots, N-1$, together with  the superposition state of the controlling qubit $K$.

Now we double those  states  $|k\rangle_{\chi}$  and  $|l\rangle_{\psi}$ that are linked to the excited state of the one-qubit subsystem $K$ and  create the same states  $|k\rangle_{\tilde \chi}$ and  $|l\rangle_{\tilde \psi}$ of the subsystems $\tilde\chi$ and $\tilde\psi$. 
  For this purpose we introduce the
projectors
\begin{eqnarray}\label{PP}
P_{\chi_i K} = |1\rangle_{\chi_i} |1\rangle_K \,  _{\chi_i}\langle 1|_K\langle 1|,\;\;
P_{\psi_i K} = |1\rangle_{\psi_i} |1\rangle_K \,  _{\psi_i}\langle 1|_K\langle 1|,\;\; i=1,\dots, n,
\end{eqnarray}
and the controlled operator
\begin{eqnarray}\label{W1}
W^{(1)}_{\chi\tilde\chi\psi\tilde\psi  K}&=& \prod_{i=1}^n \Big(P_{\chi_iK} \otimes \sigma^{(x)}_{\tilde\chi_i}  + (I_{\chi_iK} -P_{\chi_iK}) \otimes  I _{\tilde\chi_i }\Big)
 \Big(P_{\psi_iK} \otimes \sigma^{(x)}_{\tilde\psi_i}  + (I_{\psi_iK} -P_{\psi_iK}) \otimes  I _{\tilde\psi_i }\Big),
\end{eqnarray}
whose depth is $O(n)$.
Hereafter the subscript attached to the notation of a subsystem indicates the appropriate qubit of this  subsystem. Thus, the subscript $i$ in the right hand side of Eq.(\ref{W1}) denotes  the $i$th  qubit of the appropriate subsystem  $\chi$, $\tilde \chi$, $\psi$, $\tilde \psi$.
Applying  $W^{(1)}_{\chi\tilde\chi\psi\tilde\psi  K}$ to $|\Phi_1\rangle$ we obtain
\begin{eqnarray}\label{Phi2}
|\Phi_2\rangle &=& W^{(1)}_{\chi\tilde\chi\psi\tilde\psi  K} |\Phi_1\rangle = \frac{1}{2^{(2n+1)/2}}\left(
\sum_{k,l=0}^{N-1} |\Psi\rangle_{RCL}  |k\rangle_\chi|0\rangle_{\tilde\chi}  |l\rangle_\psi |0\rangle_{\tilde\psi}   |0\rangle_K\right. \\\nonumber
&+& \left.\sum_{k,l=0}^{N-1}|\Psi\rangle_{RCL}  |k\rangle_\chi|k\rangle_{\tilde\chi}  |l\rangle_\psi |l\rangle_{\tilde\psi}  |1\rangle_K)\right) + |g_2\rangle,
\end{eqnarray}
where  the garbage $|g_2\rangle$ collects all extra  terms, it is orthogonal to the first term in Eq.(\ref{Phi2}).

Next, we  prepare and apply the unitary operators  $U_\chi(\alpha)$ and $U_\psi(\beta)$ (see  Eq. (\ref{obj})) controlled  by the excited state of the  subsystem $K$ by means of the operator $W^{(2)}_{\chi\psi K }$:
\begin{eqnarray}\label{W2}
&&W^{(2)}_{\chi\psi K }=W^{(2)}_{\chi K } W^{(2)}_{\psi K },\\\label{W2r1}
&& W^{(2)}_{\chi K } = |1\rangle_K\, _K\langle 1 |\otimes U_{\chi} (\alpha) +   |0\rangle_K\, _K\langle 0 |\otimes I_{\chi}, \;\; 
W^{(2)}_{\psi K } = |1\rangle_K\, _K\langle 1 |\otimes U_{\psi} (\beta) +   |0\rangle_K\, _K\langle 0 |\otimes I_{\psi}.
\end{eqnarray}
It is convenient to represent the action of  the operators $U_\chi$ and $U_\psi$ on the  vectors $|k\rangle_\chi$ and  $|k\rangle_\psi$ in terms of  their matrix  elements  as follows:
\begin{eqnarray}\label{UU}
U_{\chi}(\alpha) |k\rangle_{\chi}  = \sum_{l_1=0}^{N-1}u_{l_1k} (\alpha)|l_1\rangle_{\chi},\;\;
U_{\psi} (\beta) |l\rangle_{\psi}  = \sum_{l_2=0}^{N-1} u_{l_2l}(\beta) |l_2\rangle_\psi.
\end{eqnarray}
The depth of the operator   $W^{(2)}_{\chi\psi K }$ will be discussed in Sec.\ref{Section:W2W2}.
Then, applying $W^{(2)}_{\chi\psi K }$ to $|\Phi_2\rangle$ we obtain

\begin{eqnarray}\label{Phi02}
|\Phi_3\rangle &=&W^{(2)}_{\chi\psi K}|\Phi_2\rangle
\\\nonumber
& =&
 \frac{1}{2^{(2n+1)/2}}\left(
\sum_{k,l=0}^{N-1} |\Psi\rangle_{RCL} |k\rangle_\chi |0\rangle_{\tilde\chi}  |l\rangle_\psi 
|0\rangle_{\tilde\psi} |0\rangle_K\right. \\\nonumber
&+& \left.\sum_{k,l=0}^{N-1}
\sum_{l_1=0}^{N-1}\sum_{l_2=0}^{N-1} u_{l_1k}(\alpha) u_{l_2l}(\beta) |\Psi\rangle_{RCL} |l_1\rangle_\chi |k\rangle_{\tilde\chi}|l_2\rangle_\psi |l\rangle_{\tilde\psi}   |1\rangle_K)\right) +|g_3\rangle.
\end{eqnarray}
Here, the garbage $|g_3\rangle$ collects  all  states that must  be removed, it  is orthogonal to the first term in (\ref{Phi02}).  
To  multiply the matrices $A$ and $B$ by the matrices $U^T_{\chi}(\alpha)\equiv U^T(\alpha)$ (from the left) and $U_{\psi}(\beta)\equiv U(\beta) $ (from the right)
and  eventually calculate the sums
 $\sum_{i,j} \,| _\psi\langle i | U^T(\alpha) A U(\beta)|j\rangle_\psi|^2$,    
 $\sum_{i,j} \,| _\psi\langle i | U^T(\alpha) B U(\beta)|j\rangle_\psi|^2$ in (\ref{obj}), we proceed according to
 Refs.\cite{ZQKW_2024, ZBQKW_2024}. Using projectors $P_{\chi_i K}$ and $P_{\psi_i K}$, given in    Eqs.(\ref{PP}),
we  introduce the following controlled operators:
 \begin{eqnarray}
 C^{(1)}_{\chi KR} &=& \prod_{i=1}^{n} \Big( P_{\chi_i K} \otimes \sigma^{(x)}_{R_i} +
 (I_{\chi_iK} -P_{\chi_iK} ) \otimes I _{R_i} \Big),\\\nonumber
  C^{(2)}_{\psi K C} &=& \prod_{i=1}^{n} \Big( P_{\psi_iK} \otimes \sigma^{(x)}_{C_i} +
 (I_{\psi_iK} -P_{\psi_iK} ) \otimes I _{C_i} \Big) ,
 \end{eqnarray}
 Here, the operator $C^{(1)}_{\chi K R}$  is required for multiplying $U^T(\alpha)$ and $ A$ ($B$), the operator
  $C^{(2)}_{\psi K C}$ serves  for multiplying $A$ ($B$) and  $U(\beta)$.
 Applying the operator $W^{(3)}_{RC\chi\psi  K} = C^{(2)}_{\psi K C} C^{(1)}_{\chi K R}$, whose depth is $O(n)$,  to the state
$ |\Phi_3\rangle$  we obtain
\begin{eqnarray}\label{Phi022}
&&|\Phi_4\rangle =W^{(3)}_ {RC\chi\psi  K}   |\Phi_3\rangle
\\\nonumber
&& = \frac{1}{2^{(2n+1)/2}}\left(
\sum_{k,l=0}^{N-1} \big(a_{00} |0\rangle_L+b_{00} |1\rangle_L\big)  |0\rangle_R|0\rangle_C |k\rangle_\chi |0\rangle_{\tilde\chi}   |l\rangle_\psi 
 |0\rangle_{\tilde\psi}  |0\rangle_K\right. \\\nonumber
&&+ \left.\sum_{k,l=0}^{N-1}
\sum_{l_1=0}^{N-1}\sum_{l_2=0}^{N-1} u_{l_1k}(\alpha) u_{l_2l}(\beta) \big(a_{l_1l_2} |0\rangle_L +b_{l_1l_2} |1\rangle_L \big) |0\rangle_R|0\rangle_C |\rangle |l_1\rangle_\chi   |k\rangle_{\tilde\chi}   |l_2\rangle_\psi |l\rangle_{\tilde\psi}  |1\rangle_K\right)+|g_4\rangle
\end{eqnarray}
where the first part in the right hand side collects the terms needed for further calculations  (these terms will be labelled later on by the operator $W^{(6)}_{ R C \chi\psi \tilde B B}$, see eq.(\ref{W5})) and $|g_4\rangle$ (orthogonal to the first term in (\ref{Phi022})) is  the garbage to be removed.
Now,  
 to complete the multiplications  $U^T(\alpha) AU(\beta)$ and $U^T(\alpha) BU(\beta)$
according to the multiplication algorithm  (see Appendix in Ref.\cite{ZBQKW_2024}), we introduce  the  operator
\begin{eqnarray}
W^{(4)}_{\chi\psi} = H_{\chi}H_\psi ,
\end{eqnarray}
where  $H_{\chi}$ and   $H_\psi$  are the sets of Hadamard transformations applied to each qubit of the subsystems $\chi$ and $\psi$ respectively, and  apply $W^{(4)}_{\chi \tilde\chi\psi}$ to the state $  |\Phi_4\rangle$. Selecting only the needed terms and moving others to the garbage $|g_5\rangle$, we obtain
\begin{eqnarray}\label{Phi023}
&&|\Phi_5\rangle =W^{(4)}_{\chi\tilde\chi\psi}|\Phi_4\rangle
\\\nonumber
&& =
 \frac{1}{2^{(4n+1)/2}}\left(\big(\tilde a_{00} |0\rangle_L+\tilde b_{00} |1\rangle_L\big)   |0\rangle_{\tilde\chi} |0\rangle_{\tilde\psi}|0\rangle_K+ \sum_{k,l=0}^{N-1} \big(A_{kl} |0\rangle_L +B_{kl} |1\rangle_L\big)  |k\rangle_{\tilde\chi} |l\rangle_{\tilde\psi} |1\rangle_K\right) |0\rangle_R|0\rangle_C |0\rangle_\chi |0\rangle_\psi +|g_5\rangle,
\end{eqnarray}
where,
\begin{eqnarray}\label{tildeAM}
&&\tilde a_{00} =2^{2n}  a_{00},\;\;\tilde b_{00} =2^{2n}  b_{00},\\\label{AM}
&&
A_{kl} (\alpha,\beta)= \sum_{l_1=0}^{N-1} \sum_{l_2=0}^{N-1}  
u_{l_1k}(\alpha) u_{l_2l}(\beta) \, a_{l_1l_2}  = (U^T(\alpha) A U(\beta))_{kl},
\\\label{AM2}
&&
B_{kl} (\alpha,\beta)= \sum_{l_1=0}^{N-1} \sum_{l_2=0}^{N-1}  u_{l_1k}(\alpha) u_{l_2l}(\beta) \, b_{l_1l_2} = (U^T(\alpha) B U(\beta))_{kl}.
\end{eqnarray}
We note that the factor $2^{2n}$ in  the expressions for  $\tilde a_{00}$  and  $\tilde b_{00}$  in (\ref{tildeAM}) appears  because of the sum over $k$ and $l$ in the first term in the parenthesis of (\ref{Phi022}) which includes $N^2$ terms.

Now we label those terms in the sum in Eq.(\ref{Phi023})  that are needed for loss function  according to the sum in Eq.(\ref{obj}). 
To this end, we introduce the projectors
\begin{eqnarray}
&&p_{jk} = |j\rangle _{\tilde\chi}  |k\rangle _{\tilde\psi}  |1\rangle_K \; _{\tilde\chi}\langle j| _{\tilde\psi}\langle k| _K\langle 1 |,\;\; j=1,\dots, N-1,\;\; k=0,\dots, j-1,
\end{eqnarray}
the one-qubit ancilla $\tilde B$ in the ground state and the controlled operator
\begin{eqnarray}
&&
W^{(5)}_{\tilde \chi \tilde \psi K \tilde B} = 
P_{\tilde \chi \tilde \psi K} \otimes \sigma^{(x)}_{\tilde B} + (I_{\tilde \chi \tilde \psi K}-P_{\tilde \chi \tilde \psi K} )\otimes I_{\tilde B},\\\nonumber
&&
P_{\tilde \chi \tilde \psi K} =
\sum_{j=1}^{N-1} \sum_{k=0}^{j-1} p_{jk},
\end{eqnarray}
whose depth is $O(N^{2}\log N)$.
Applying $W^{(5)}_{\tilde \chi \tilde \psi K \tilde B}$ to $|\Phi_5\rangle |0\rangle_{\tilde B}$ we obtain
\begin{eqnarray}\label{Phi6}
&&
|\Phi_6\rangle =W^{(5)}_{\tilde \chi \tilde \psi K \tilde B} |\Phi_5\rangle |0\rangle_{\tilde B}
\\\nonumber
& &=
 \frac{1}{2^{(4n+1)/2}}\left(\big(\tilde a_{00} |0\rangle_L+\tilde b_{00} |1\rangle_L\big)  |0\rangle_{\tilde\chi} |0\rangle_{\tilde\psi} |0\rangle_K+ \sum_{k=1}^{N-1} \sum_{l=0}^{k-1} \big(A_{kl} |0\rangle_L +B_{kl} |1\rangle_L\big) |k\rangle_{\tilde\chi} |l\rangle_{\tilde\psi} |1\rangle_K\right) \times \\\nonumber
 && |0\rangle_R|0\rangle_C |0\rangle_\chi |0\rangle_\psi  |1\rangle_{\tilde B}+|g_6\rangle ,
\end{eqnarray}

Now we label and remove the garbage $|g_5\rangle$ from the state (\ref{Phi6})   via the measurement.        To this end, we
introduce another one-qubit ancillae $B$  in the ground state and the controlled operator
\begin{eqnarray}\label{W5}
&&
W^{(6)}_{ R C \chi \psi \tilde B B} =P\otimes \sigma^{(x)}_B +
(I_{R C\chi \psi  \tilde B}-P)\otimes I_{B},\\\nonumber
&&
P= |0\rangle_R|0\rangle_C |0\rangle_{\chi}  |0\rangle_\psi |1\rangle_{\tilde B}\,
 _R\langle 0| _C\langle 0|  _{\chi}\langle 0|  _\psi\langle 0|  _{\tilde B}\langle 1| ,
\end{eqnarray}
whose depth is $O(\log N)$.
Then, applying $W^{(6)}_{ R C \chi \psi  \tilde B B}$ to the state $ |\Phi_6\rangle |0\rangle_B$ we obtain
\begin{eqnarray}\label{Phi3}
&&
|\Phi_7\rangle = W^{(6)}_{R C\chi\psi \tilde B B} |\Phi_6\rangle |0\rangle_B=
\\\nonumber
&&
   \frac{1}{2^{(4n+1)/2}}\left( \big(\tilde a_{00} |0\rangle_L+\tilde b_{00} |1\rangle_L\big)   |0\rangle_{\tilde\chi} |0\rangle_{\tilde\psi} |0\rangle_K+ \sum_{k=1}^{N-1} \sum_{l=0}^{k-1}  \big(A_{kl} |0\rangle_L +B_{kl} |1\rangle_L \big)  |k\rangle_{\tilde\chi} |l\rangle_{\tilde\psi} |1\rangle_K\right) \times \\\nonumber
 &&|0\rangle_R|0\rangle_C |0\rangle_\chi |0\rangle_\psi |1\rangle_{\tilde B}  |1\rangle_B+|g_6\rangle |0\rangle_B .
\end{eqnarray}
Now we can remove the garbage by  applying   the  measurement operator $M_B$ 
 to the ancilla $B$. By measurement operator $M_B$ we mean the operator that, been applied to the superposition 
 state $\alpha|0\rangle_B + \beta |1\rangle_B$, $|\alpha|^2 + |\beta|^2=1$, yields either
  $ |0\rangle_B$  or $|1\rangle_B$ with probability, respectively,  $|\alpha|^2$ or $|\beta|^2$.   Thus, applying $M_B$ to $B$    with the desired output $|1\rangle_B$, whose probability (success probability) is  
 \begin{eqnarray}\label{sp}
 p=\frac{G^2}{2^{4n+1}},\;\; G=\sqrt{ |\tilde a_{00}|^2+|\tilde b_{00}|^2+ \sum_{k=1}^{N-1}  \sum_{l=0}^{k-1}(|A_{kl}|^2+|B_{kl}|^2)} =\sqrt{|\tilde a_{00}|^2+|\tilde b_{00}|^2 + L},
 \end{eqnarray}
 we reduce the  state of the remaining system to the  following one:
\begin{eqnarray}\label{Phi4}
&&
|\Phi_8\rangle =
|\Psi_{out}\rangle |0\rangle_R |0\rangle_C |0\rangle_{\chi}
   |0\rangle_{\psi}  |1\rangle_{\tilde B} |1\rangle_B ,
   \\\nonumber
   &&
|\Psi_{out}\rangle=   G^{-1}\left(\big(\tilde a_{00} |0\rangle_L+\tilde b_{00} |1\rangle_L\big)    |0\rangle_{\tilde\chi} |0\rangle_{\tilde\psi} |0\rangle_K+ \sum_{k=1}^{N-1} \sum_{l=0}^{k-1} \big(A_{kl} |0\rangle_L +B_{kl} |1\rangle_L \big) |k\rangle_{\tilde\chi} |l\rangle_{\tilde\psi} |1\rangle_K\right).
\end{eqnarray}
Now, the loss function $L$ and normalization $G$ can be probabilistically found by measuring the state of $K$.
The probability of obtaining $|0\rangle_K$ and $|1\rangle_K$  equal, respectively
\begin{eqnarray}
p_0=\frac{ |\tilde a_{00}|^2+|\tilde b_{00}|^2}{G^2},\;\; p_1=\frac{L}{G^2}, \;\; p_0+p_1=1.
\end{eqnarray}
Thus, we have
\begin{eqnarray}\label{Lp}
L=\big(|\tilde a_{00}|^2+|\tilde b_{00}|^2\big)\frac{p_1}{1-p_1},\;\;  G^2 = \frac{|\tilde a_{00}|^2+|\tilde b_{00}|^2}{1-p_1}.
\end{eqnarray}
This is the last step of the quantum part of the hybrid algorithm for calculating GEV. Now we give details regarding the representation of the unitary operators $U_\chi(\alpha)$ and $U_\psi(\beta)$, given in Eqs.(\ref{UU}),  in terms of  one-qubit rotations and C-NOTs.

\noindent

\subsubsection{Realization of  operator $W^{(2)}_{\chi\psi K}$}
\label{Section:W2W2}
Constructing the unitary operations  $U_\chi$ and $U_\psi$ included into the operator $W^{(2)}_{\chi\psi K}$ in Eq. (\ref{W2})  we follow Refs. \cite{ZQW_2025,WSW}.
The operator $W^{(2)}_{\chi\psi K}$
is the superposition of  two operators $W^{(2)}_{\chi K }$ and $W^{(2)}_{\psi K }$  which  are completely equivalent to each other  and defer only by the parameters encoded into them.
 Therefore, we describe only
$W^{(2)}_{\chi K}$.
 To realize the operator $U_\chi$ for triangularization of the complex  matrices  $A$ and $B$,  it is enough to use the  one-qubit  operators
$r_{\chi_j}(\alpha_l) =R_{z}(\alpha^{(1)}_l) R_{y}(\alpha^{(2)}_l)  R_{z}(\alpha^{(3)}_l)$, where
$R_{z}(\alpha^{(k)}_l)= \exp(- i \sigma^{(z)} \alpha^{(k)}_l/2)$, $k=1,3$, $R_{y}(\alpha^{(2)}_l) =\exp(- i \sigma^{(y)} \alpha^{(2)}_l/2)$ { ($\sigma^{(y)}$  and $\sigma^{(z)}$ are the Pauli matrices)}, and C-nots \cite{WSW}:
\begin{eqnarray}\label{UU2}
&&
U_\chi(\alpha) = \prod_{k=1}^M R_k(\alpha_{(k-1)n +1},\dots,\alpha_{kn}) , \;\; \alpha_j =\{ \alpha_j^{(1)}, \alpha_j^{(2)}, \alpha_j^{(3)}\},\;\;\alpha=\{\alpha_j: j=1,\dots, M n\},\\\nonumber
&&
R_k(\alpha_{(k-1)n +1},\dots,\alpha_{kn})   = \prod_{m=1}^{n-1} C_{\chi_m\chi_{m+1}}  \prod_{j=1}^{n} r_{\chi_j}(\alpha_{(k-1)n +j}) ,\\\nonumber
&&C_{\chi_m\chi_{m+1}} = |1\rangle_{\chi_m}\,_{\chi_m}\langle 1| \otimes \sigma^{(x)}_{\chi_{m+1}} +  |0\rangle_{\chi_m}\,_{\chi_m}\langle0 | \otimes I_{\chi_{m+1}} .
\end{eqnarray}
In this formula, $R_k$ represents a single block of transformations encoding $3 \log N$ parameters $\alpha^{(k)}_j$, $k=1,2,3$, $j=1,\dots, \log N$. Involving $M$ blocks $R_k$, $k=1,\dots, M$, we
enlarge the number of parameters to $3 M\log N$. We note that the  number of free real parameters in the
$N\times N$ unitary transformation  is $N^2$. Therefore, $M\ge \frac{N^2}{3\log N}$ to reproduce the most general unitary transformation.
The realization of unitary transformation in terms of rotation operators   is chosen for two reasons: (i)
simple realization of $U$  in terms of one- and two-qubit operations and (ii) simple calculation  of  derivatives of the
loss function with respect to the optimization  parameters, see Eq.(\ref{derivative}). 

Now we turn to realization of the controlled operator $W^{(2)}_{\chi K }$  given in (\ref{W2r1}) . To this end we  use  $U_\chi$ determined in  (\ref{UU2})  in the controlled operator $W^{(2)}_{\chi K }$,  given in Eq.(\ref{W2r1}) in terms of the operators C-not, and rewrite it in the form 
\begin{eqnarray}\label{CWW}
&&
W^{(2)}_{\chi K } =\prod_{k=1}^M \prod_{m=1}^{n-1} C_{K\chi_m\chi_{m+1}} \prod_{j=1}^n
R_{z\chi_j}(\alpha^{(1)}_{(k-1)n+j}/2) C^{(x)}_{K\chi_j}R_{z\chi_j}(\alpha^{(1)}_{(k-1)n+j}/2)    \times \\\nonumber
&&
R_{y\chi_j}(\alpha^{(2)}_{(k-1)n+j}/2)    C^{(x)}_{K\chi_j}R_{y\chi_j}(\alpha^{(2)}_{(k-1)n+j}/2)     C^{(x)}_{K\chi_j}
R_{z\chi_j}(\alpha^{(3)}_{(k-1)n+j}/2) C^{(x)}_{K\chi_j}R_{z\chi_j}(\alpha^{(3)}_{(k-1)n+j}/2)    ,
\end{eqnarray}
where
\begin{eqnarray}
&&
C_{K\chi_m\chi_{m+1}}= P_{K\chi_m}\otimes \sigma^{(x)}_{\chi_{m+1}}+
( I_{K\chi_m}- P_{K\chi_m})\otimes I_{\chi_{m+1}},\\\nonumber
&&P_{K\chi_m}= |1\rangle_K |1\rangle_{\chi_m}\, _K\langle 1| _{\chi_m}\langle 1|
,
\\\nonumber
&&
C^{(x)}_{K\chi_j}= |0\rangle_K \, _K\langle 0|  \otimes \sigma^{(x)}_{\chi_j}+ |1\rangle_K \, _K\langle 1|
\otimes I_{\chi_j}.
\end{eqnarray}
It is simple to check that this operator  acts as  $I_\chi$ or  $U_\chi$ on the basis  states  of the subsystem $\chi \cup K$ including, respectively,  the state  $|0\rangle_K$ or $|1\rangle_K$ of  the subsystem  $K$.  The depth of this operator is $O(M\log N)\ge  O(N^2)$.

\subsection{Derivatives of loss function and iterations of optimization parameters}
\label{Section:der}

The input data for   optimization algorithms include the derivatives of the loss functions with respect to the optimization parameters  $\alpha=\{\alpha^{(j)}_k\}$ and $\beta=\{\beta^{(j)}_k\}$.
Since these parameters in our algorithm are introduced through the exponent of Pauli matrices $\sigma^{(y)}$ and $\sigma^{(z)}$ in Eq.(\ref{CWW}), such derivatives can be expressed in terms of the loss function with shifted values of parameters which can be calculated using the  quantum algorithm described in Sec.\ref{Section:general}. Then, the whole set of calculated loss functions can be  supplied to  the  classical algorithm which  calculates the subsequent  values of the optimization parameters for the next iteration step of the optimization algorithm.  

Let $\hat \gamma=\{\hat \gamma_1,\dots,\hat \gamma_{6nM}\}$ be the set of all parameters $\alpha$ and $\beta$: $\hat\gamma=\{\alpha,\beta\}$. Then \cite{LFYYFZDM}
\begin{eqnarray}\label{derivative}
&&
\frac{\partial L(\hat \gamma)}{\partial \hat\gamma_k} = \frac{1}{2}  
\big(L(\hat\gamma^{(k,+)})   -L(\hat\gamma^{(k,-)})\big), \;\; k=1,\dots,6nM,\\\nonumber
\end{eqnarray}
where $\hat\gamma^{(k\pm)}$ means the set of parameters $\hat\gamma$, in which $\hat\gamma_k$ is replaced with
$\hat\gamma_k\pm\pi/2$ i.e.,
\begin{eqnarray}\label{gampm}
&&
\hat\gamma^{(k,\pm)} =\hat \gamma|_{\hat\gamma_k \to \hat\gamma_k\pm\pi/2},\;\;k=1,\dots,6nM.
\end{eqnarray}
Thus, we need to calculate the loss functions $L(\hat\gamma$, $L(\hat\gamma^{(k,\pm)})$,   $k=1,\dots,6nM$, i.e., $12 n M+1$ values all together. 

After calculating derivatives (\ref{derivative}), we can calculate the new values of the parameters $\hat \gamma$:
\begin{eqnarray}\label{newgamma}
\hat\gamma_k \to \hat\gamma_k - \delta \frac{\partial L(\hat \gamma)}{\partial \hat\gamma_k} , \;\; k=1,\dots, 6 n M,
\end{eqnarray}
where $\delta$ is some small parameter, and calculate subsequent set of  $12 n M+1$ loss functions for this new set of optimization parameters and so on. The iteration stops when the algorithm reaches the required accuracy of calculating either  eigenvalues or loss function. The relation between these two accuracies  will be shown in Sec.\ref{Section:var}.

\section{Features of algorithm for calculating EV}
\label{Section:ev}
Formally, the algorithm for problem on eigenvalues  of  the matrix $A$ can be deduced from the described algorithm putting $B=I$ in  Eq.(\ref{obj}). However, in this case it is natural  to use the triangularization formula \cite{GL}
\begin{eqnarray}
Q^\dagger AQ=T, \\\label{lambdakEV}
 \lambda_k = t_{kk},
\end{eqnarray}
where $Q$ is the unitary matrix and $T$ is the upper-triangular matrix, $\lambda_k$  and $t_{kk}$ are, respectively, the eigenvalue of $A$ and the diagonal element of $T$. 
For this case, the loss function (\ref{obj}) should be replaced with 
\begin{eqnarray}\label{objev}
&&
L(\beta) = \sum_{j=1}^{N-1}  \sum_{j=0}^{i-1}   |\langle i| U^\dagger (\beta) A U(\beta)|j\rangle|^2  =
 \sum_{j=1}^{N-1}  \sum_{j=0}^{i-1}  |\langle i| T(\beta)|j\rangle|^2,
\end{eqnarray}
reducing twice the number of optimization parameters. Since there is no matrix $B$, we do not need the one-qubit subsystem $L$, the circuit in Fig.\ref{Fig:SVD} must be modified by removing the subsystem  $L$ and  replacing $U(\alpha)\to U^*(\beta)$, where $*$  denotes the complex conjugate.
Now  the number of parameters in the list $\hat\gamma$ is $3 n M$: $\hat \gamma=\{\hat \gamma_1,\dots,\hat \gamma_{3nM}\}=\{\beta^{(n)}_j: n=1,2,3, j=1,\dots nM\}$. 
 Since all the steps of the algorithm with such simplified loss function are exactly the same, we do not give more details of  the quantum algorithm for calculating the loss function. If the optimizing values $\beta_0$ of the $\beta$-parameters are found,  then $Q=U(\beta_0)$.

However, the formula (\ref{derivative}) for the derivatives of the loss function must be modified. 
Since each parameter from the list  $\hat\gamma$ appears four times in the expression for the loss function (\ref{objev}), the formula for the derivative of the loss function with respect to the optimization parameters becomes as follows:
\begin{eqnarray}\label{derivative2}
&&
\frac{\partial L(\hat \gamma)}{\partial \hat\gamma_k} =-\frac{\sin^2\alpha/2}{2 \cos \alpha}\left( 2 \Big(L(\hat\gamma^{k,+}) -L(\hat\gamma^{k,-}) \Big) -\frac{1}{ \sin^2 \alpha/2  \sin  \alpha} \Big(L(\hat\gamma^{k,+\alpha}) -L(\hat\gamma^{k,-\alpha})\Big)    \right)
  ,
\end{eqnarray}
where $\hat\gamma^{(k,\pm)}$ are defined in (\ref{gampm}),   and   $\hat\gamma^{(k,\pm\alpha)}$ denotes  the set of parameters $\hat\gamma$, in which $\hat\gamma_k$ is replaced with
$\hat\gamma_k\pm \alpha$, $\alpha\ge 0$, $\alpha \neq n \pi/2$, $n\in \mathbb Z$.
Thus, we calculate the set of  loss functions including $L$, $L(\hat\gamma^{k,\pm})$, $L(\hat\gamma^{k,\pm\alpha})$,
$k=1,\dots,3 n M$, i.e., $12n M+1$ values all together, the same as in Sec.\ref{Section:SVD}. Then, we proceed to calculating the derivatives of the loss function by using Eq.(\ref{derivative2}), constructing subsequent values of the optimization parameters by using Eq.(\ref{newgamma}), calculate the subsequent set of  $12 n M+1$ loss functions for the  new set $\hat \gamma$ by using the quantum algorithm  and so on. The iteration process stops when the required accuracy of calculation is achieved. 

\section{Characteristics of algorithms}
\label{Section:var}
Both algorithms considered in  Secs.\ref{Section:SVD} and \ref{Section:ev} have similar  parameters characterizing their effectiveness. 

1.{\it The  depth}   of the circuit is defined by the operators
   $W^{(2)}_{\chi\psi K}$ and
  $W^{(5)}_{\tilde \chi \tilde \psi K \tilde B}$ whose depths are
 $O(M \log N) $ and
   $O(N^2 \log N)$ respectively.  Depending on $M$, either of two depth can prevail.
   
2. {\it The space} required for realization of quantum  algorithm  is $O(\log N)$ qubits. 

4. {\it The success probability $p$},  $p\sim N^{-4}$ according to  Eq.(\ref{sp}), is the probability of removing the garbage via the measurement of the state of the ancilla $B$.  

3. {\it The accuracy $\varepsilon$ of calculating the eigenvalues} is the principal parameter characterizing the accuracy of calculation. It is related to other accuracies listed below.

4. {\it The number of iterations  $N^{(it)}$} needed to obtain the eigenvalues  with required accuracy $\varepsilon$.  The numerical simulations in Sec. \ref{Section:num} show that, at least for the considered example,   $N^{(it)}= -\varkappa_0 \log C_0 \varepsilon$, where the parameters $\varkappa_0$ and $C_0$ are defined by the matrices $A$ and $B$ used in generalized eigenvalue problem, by the parameter $M$ in the unitary transformations $U_\chi$, $U_\psi$ and by particular gradient method implemented in the algorithm.

5. {\it The accuracy $\varepsilon_L$ of calculating  the loss function}  characterizes the deviation of the optimized $L$ from zero: $L\le \varepsilon_L$.

6. {\it The number of runs of the algorithm.} To probabilistically find the loss function $L(\hat \gamma)$ and  all  $L(\hat\gamma^{(k)})$ 
 needed for calculating the derivatives at fixed values of  the parameters, we have to perform $N^{(r)}_1 = 12n M+1$  runs of the algorithm, as shown in Secs. \ref{Section:der},  \ref{Section:ev}. 
In addition, each  iteration  requires $N^{(r)}_2\sim \frac{1}{\varepsilon_p}$ runs to find probability $p_1$ in Eq.(\ref{Lp}) with the accuracy $\varepsilon_p$ ($p_1=p_1^{(0)} +\varepsilon_p$). The total number of runs is $N^{(r)}=N^{(r)}_1 N^{(r)}_2$.
 In turn, $\varepsilon_p$  is related to the accuracy    $\varepsilon_L$  required for  probabilistic  calculation of  the  loss function $L$. This relation follows from Eq.(\ref{Lp}):
\begin{eqnarray}
\label{epsilonL}
&&
L = a\frac{p_1}{p_0} = a \frac{p_1^{(0)} +\varepsilon_p}{1-p_1^{(0)}- \varepsilon_p}  \approx 
\frac{a p_1^{(0)}}{1-p_1^{(0)}} + \Delta L, \\\nonumber
&&
a=\big(|\tilde a_{00}|^2+|\tilde b_{00}|^2\big),\;\;  \Delta L =  \frac{a}{(1-p_1^{(0)})^2} \varepsilon_p,
\end{eqnarray} 
where we assume that $a\neq 0$. 
Obviously, to provide the accuracy $\varepsilon_L$, we require
\begin{eqnarray}
\Delta L \lesssim \varepsilon_L .
\end{eqnarray}

8. {\it The accuracy $\sigma$ of encoding the parameters $\hat \gamma$.} The optimization parameters $\hat\gamma$ in the unitary transformations $U_\chi$ and $U_\psi$ 
can not be realize with absolute accuracy. We introduce the parameter $\sigma=10^{-d}$, where $d$ is the 
number of decimals kept in the parameters $\hat\gamma$, thus 
$\hat\gamma_k =\hat  \gamma_k^{(0)} + \hat \gamma_k^{(\sigma)} \sigma$, where
 $\hat \gamma_k^{(0)}$ is 
the encoded  parameter and $\hat \gamma_k^{(\sigma)}$ characterizes the neglected part of the parameter 
$\hat \gamma_k$.

\vspace{0.5cm}


The  accuracies  of calculating  the eigenvalues and loss function, parameters $\varepsilon$ and $\varepsilon_L$, are used in the classical part of the algorithm.  These two parameters can be related to each other as follows. 
Let the matrix elements of $T$ and $S$ in the formula for the loss function, Eq. (\ref{obj}),  be  found with the accuracy $\tilde \sigma \ll 1$, i.e., $t_{ij}=t_{ij}^{(0)}+ t^{(\tilde \sigma)}_{ij} \tilde \sigma $, $s_{ij}=s_{ij}^{0)}+ s^{(\tilde \sigma)}_{ij}\tilde \sigma$, where $t^{(0)}_{ij}$ and   $s^{(0)}_{ij}$ are some  mean values of the appropriate matrix elements, while $t^{(\tilde \sigma)}_{ij}$ and   $s^{(\tilde \sigma)}_{ij}$ characterize deviations from the mean values.  The loss function $L$ includes only vanishing elements of  $T$ and $S$ for which the mean values are zero, therefore 
\begin{eqnarray}\label{Le}
L = L^{(\tilde \sigma)} \tilde \sigma^2 = \varepsilon_L,
\end{eqnarray}
 where  $L^{(\tilde \sigma)}>0 $ can be written in terms of $t^{(\tilde \sigma)}_{ij}$ and $s^{(\tilde \sigma)}_{ij}$ (we do not present exact formula). The accuracy $\varepsilon$ for  the eigenvalues can be expressed in terms of $\tilde\sigma$ following Eq.(\ref{lambdak}) (we consider the accuracy for GEV; the accuracy  for  EV can be obtained similarly using  Eq.(\ref{lambdakEV})):
\begin{eqnarray}\label{le}
\lambda_k=\lambda^{(0)}_k + \lambda^{(\tilde \sigma)}_k \tilde \sigma,\;\;
\lambda^{(0)}_k =\frac{t^{(0)}_{kk}}{s^{(0)}_{kk}},  \;\;\lambda^{(\tilde \sigma)}_k=  \left( \frac{t^{(\tilde \sigma)}_{kk}}{s^{(0)}_{kk}}  -
 \frac{t^{(0)}_{kk} s^{(\tilde \sigma)}_{kk} }{(s^{(0)}_{kk})^2}\right),
 \end{eqnarray} 
 where $\lambda^{(0)}_k$ is the exact GEV; $\lambda^{(\tilde \sigma)}_k=t^{(\tilde \sigma)}_{kk}$ in the case of Eq.(\ref{lambdakEV}) for EV.
 We can introduce  $\varepsilon$ as follows:
 \begin{eqnarray}\label{le2}
 \varepsilon =\lambda^{(\tilde \sigma)} \tilde \sigma, \;\;\lambda^{(\tilde \sigma)}=\frac{\tilde \sigma}{N}  \sum _{k=1}^N |\lambda^{(\tilde \sigma)}_k| .
 \end{eqnarray}
Comparison of (\ref{Le}) and (\ref{le2}) yields
\begin{eqnarray}\label{ee}
\varepsilon_L = \frac{L^{(\tilde \sigma)}}{(\lambda^{(\tilde \sigma)})^2} \varepsilon^2. 
\end{eqnarray}
The factor at  $\varepsilon^2$ in the right hand side of (\ref{ee})  depends on a particular matrix, while the quadratic relation between 
$\varepsilon_L$ and $\varepsilon$ is universal. 

The accuracy $\sigma$ of encoding the parameters $\hat \gamma$ yields  the lower boundary   $\varepsilon^{(min)}$ for the  accuracy of 
calculation of GEV and EV.  
To estimate this lower boundary, we note that $\sigma$-accuracy for $\hat \gamma$ results in perturbation of the  operators $U_\chi$ and $U_\psi$:
$
U_\chi(\alpha) = U_\chi(\alpha^{(0)}) +   \sum_{j>0} U_\chi^{(j)} \sigma^j
$ and similar for $U_\psi(\beta)$.
 Substituting these perturbed  unitary transformations into the loss function (\ref{obj})  we obtain, up to the first order in $\sigma$,
 \begin{eqnarray}\label{ee2}
 L = L(\gamma^{(0)}) +L^{(\sigma)} \sigma,
 \end{eqnarray}
we do not present the expression for $L^{(\sigma)} $.  Therefore, the minimal values of  $\varepsilon_L$ is  a linear function of $\sigma$: $\varepsilon^{(min)}_L = L^{(\sigma)} \sigma$.
Then Eq.(\ref{ee}) yields: 
\begin{eqnarray}\label{ee3}
\varepsilon^{(min)} = \lambda^{(\sigma)} \sqrt{\sigma}, \;\; \lambda^{(\sigma)} = \lambda^{(\tilde \sigma)}\sqrt{\frac{L^{(\sigma)}}{L^{(\tilde \sigma)}}}.
\end{eqnarray}
Relations (\ref{ee})-(\ref{ee3}) are confirmed  by numerical simulations in Sec.\ref{Section:num}.

\section{Numerical simulations}
\label{Section:num}
We  consider the  problems on both GEV and  EV for the $4\times 4$ matrices $A$ and $B$  encoded into the states of the five-qubit subsystem $R\cup C \cup L$, where $R$ and $C$ are two-qubit subsystems ($n=2$). 
We randomly generate $100$ pairs of the matrices $A$ and $B$ under condition that $\min_i| \lambda_i (A,B)| \ge 0.1$. 
Initially, we chose the parameter $\delta=0.5$ in Eq.(\ref{newgamma}) for the iterated parameters $\hat\gamma$. Then this parameter  dynamically increases with the decrease in the value of the  loss function according to the following rule: 
\begin{eqnarray}\label{rule}
&&
 {\mbox{if}}\;\; 
\frac{2 (L^{(n-1)} - L^{(n)})}{  L^{(n-1)} +L^{(n)}} < 0.001, \;\;{\mbox{then}} \;\;  \delta \to  1.0005\, \delta;\\\nonumber
&&
{\mbox{if}} \;\; L^{(n)} >L^{(n-1)}, \;\; {\mbox{then}} \;\;  \delta \to\frac{\delta}{1.0005},
\end{eqnarray}
where $L^{(n)}$ is the value of the loss function at the $n$th iteration step.
  We set  $M=10$ in Eq.(\ref{CWW}), $\alpha =\pi/3$ in (\ref{derivative2}) ,  and  all optimization parameters $\hat \gamma$ equal to  zero initially. 

\subsection{Unperturbed optimization parameters $\hat \gamma$}
\label{Section:unpert}
\begin{figure}[ht]
\begin{subfigure}{0.45\textwidth}
    \includegraphics[width=0.7\textwidth]{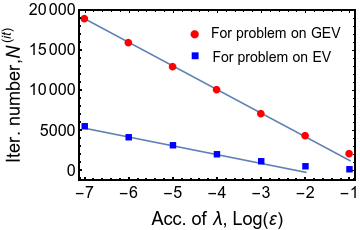}
    \caption{}
    \end{subfigure}
    \begin{subfigure}{0.45\textwidth}
      \includegraphics[width=0.7\textwidth]{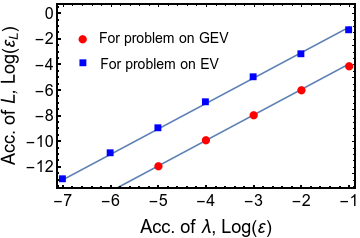}
      \caption{}
    \end{subfigure} 
    \caption{(a) The iteration number $N_r$  and  (b) the accuracy of the loss function $\varepsilon_L$ as functions of the accuracy $\log \varepsilon$ of  constructed GEV and EV.  These figures show that both $N_r$ and   $\log \varepsilon_L$ are proportional to  $\log \varepsilon$ for small enough $\varepsilon$. The red circle and blue squares  correspond, respectively, to the problem on GEV discussed in Sec.\ref{Section:SVD},  and to the problem on EV   discussed in Sec.\ref{Section:ev}.  Slopes of both lines in plate (b) are very close to each other. The algorithm for constructing EV converges  faster (iteration number is less)}
\label{Fig:LossF}
\end{figure}

In this section, we assume that the optimization parameters $\hat \gamma$ at each iteration step  can be perfectly encoded into the unitary transformations $U_\chi$ and $U_\psi$. 
The  optimization algorithm stops  iterating if  $1/4 \sum_{i=1}^4 |\lambda_i-\lambda_i^{(0)}| < \varepsilon$,  where $\lambda^{(0)}_i$ are the (generalized) eigenvalues calculated by the  classical algorithm.
We  obtain the  iteration number $N^{it}$ and logarithm of the  loss function $L =\varepsilon_L$,  averaged over experiments with $100$ pairs of random $4\times 4$ matrices  $\{A,B\}$,  as functions of $\log \varepsilon$ for  $\varepsilon =10^{-j}$, $j=1,\dots,7$. 
The averaged results are shown by the red dots  on Fig. \ref{Fig:LossF}. In this section, $\log$ means the logarithm to the base 10. 
For the family of the above 100 numerical experiments and for small enough $\varepsilon$ ($\varepsilon \le 10^{-3}$ in the considered example),  $N^{(it)}$ can be approximated by the linear function 
\begin{eqnarray}\label{Nline}
&&
N^{(it)}=  -\varkappa^{(N)} \log \varepsilon  -C^{(N)}, 
\end{eqnarray}
where $\varkappa_N $ and $C_N$ can be written as follows:
\begin{eqnarray}\label{NUM:kap}
\varkappa^{(N)}\approx\varkappa^{(N;0)} \pm \Delta \varkappa^{(N)}, \;\; C^{(N)}\approx  C^{(N;0)} \pm \Delta C^{(N)}.
\end{eqnarray}
Here the superscript $0$ denotes the mean value over $100$ experiments and $\Delta$ denotes the mean-square deviation of the appropriate parameter, for instance, 
\begin{eqnarray}\label{NUM:disp}
\varkappa^{(N;0)} =\frac{1}{100} \sum_{j=1}^{100} \varkappa^{(N)}_j,\;\;  \Delta \varkappa^{(N)} = \sqrt{\frac{1}{100} \sum_{j=1}^{100} (\varkappa^{(N)}_j - \varkappa^{(N;0)})^2},
\end{eqnarray}
where the parameter $\varkappa^{(N)}_j$ is  associated with the $j$th pair of the matrices $\{A,B\}$.
In our example
\begin{eqnarray}\label{Nex}
{\mbox{for problem on  GEV: }} &&\varkappa^{(N;0)}  =2936.66,\;\;  C^{(N;0)}=1670.77,\;\; \Delta \varkappa^{(N)} = 2229.85,\;\; 
 \Delta C^{(N)} =  4257.67,\\\nonumber
{\mbox{for problem on EV: }} && \varkappa^{(N;0)}  = 1094.51,\;\;  C^{(N;0)}=2373.77,\;\; \Delta \varkappa^{(N)} = 1882.98,\;\; 
 \Delta C^{(N)} = 5939.05.
\end{eqnarray}
The functions $N^{(it;0)} = -\varkappa^{(N;0)} \log \varepsilon  -C^{(N;0)}$  for the both algorithms are shown in  Fig.\ref{Fig:LossF}a by the lines. The red circles and blue squares on this  graph well fit the appropriate  lines for $\varepsilon \le10^{-3}$.
The large mean-square deviation of both parameters $\varkappa^{(N)}$ and $C^{(N)}$ means that the lines approximating dependence $N^{(it)}(\varepsilon)$ are completely different for different pairs of matrices $\{A,B\}$. 
In addition, the  iteration number $N^{(it)}$    depends on the parameter $M$ in the unitary transformation and decreases with an increase in $M$ (till some saturation value), but the increase in $M$ means the  increase in the depth of the algorithm, therefore one has to manipulate between $M$ and $N^{(it)}$ to get optimal evaluation time for the algorithm.  

Similarly, the parameter $\log \varepsilon_L$ as   a function of $\log \varepsilon$  for the considered  family of 100 numerical experiments can be also approximated by the linear function for $\varepsilon \le 10^{-3}$:
 \begin{eqnarray}\label{Lappr}
 &&
 \log \varepsilon_L = \varkappa^{(L)} \log \varepsilon -C^{(L)}\;\;\Rightarrow \;\;  \varepsilon_L = 10^{- C^{(L)}} \varepsilon^{\varkappa^{(L)}},\\\nonumber
 && \varkappa^{(L)}=\varkappa^{(L;0)}\pm \Delta \varkappa^{(L)}, \;\; C^{(L)}=C^{(L;0)}\pm \Delta C^{(L)}, \\\nonumber
{\mbox{for  problem on GEV: }}  &&
  \varkappa^{(L;0)}= 1.990, \;\; C^{(L;0)}= 1.914, \;\; \Delta \varkappa^{(L)} =   0.088,\;\; \Delta C^{(L)}=  4.041, \\\nonumber
{\mbox{for problem on EV: }}  &&
  \varkappa^{(L;0)}= 1.986, \;\; C^{(L;0)}=0.953, \;\; \Delta \varkappa^{(L)} =  0.071,\;\; \Delta C^{(L)}=   1.203,
 \end{eqnarray}
 where the superscript $0$ and $\Delta$ mean the same as in Eqs. (\ref{NUM:kap}), see definitions in Eqs. (\ref{NUM:disp}). We emphasize that expression (\ref{Lappr}) for $\varepsilon_L$ agrees with  the quadratic dependence of $\varepsilon_L$ on $\varepsilon$ in Eq. (\ref{ee}) because $\varkappa^{(L)} \approx 2$ in this case; the mean-square deviation $\Delta \varkappa^{(L)}$ is $\approx 4\%$ of $\varkappa^{(L;0)}$ for  both problems on GEV and on EV (i.e. the slope of all 100 lines in Eq.(\ref{Lappr})  is almost the same), unlike the mean-square deviation of the parameter $\varkappa^{(N)}$  in line (\ref{Nline})  approximating  $N^{(it)}$, see Eqs.(\ref{Nex}).  On the contrary, the mean-square deviation of 
 $C^{(L)}$ is large (it exceeds the mean values of  $C^{(L)}$), i.e., this parameter significantly depends on the considered matrices. We also emphasize that the slopes of lines for  problems on  GEV and EV on Fig.\ref{Fig:LossF}b are very close to each other, which confirms  Eq.(\ref{ee}) that holds for both algorithms.
 
\subsection{Perturbation of optimization parameters $\hat\gamma$}

\begin{figure}[ht]
\begin{subfigure}{0.4\textwidth}
    \includegraphics[width=0.7\textwidth]{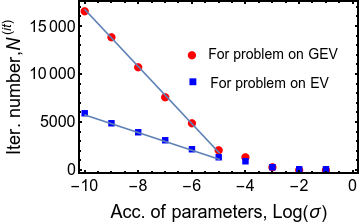}
    \caption{}
    \end{subfigure}
    \begin{subfigure}{0.4\textwidth}
      \includegraphics[width=0.7\textwidth]{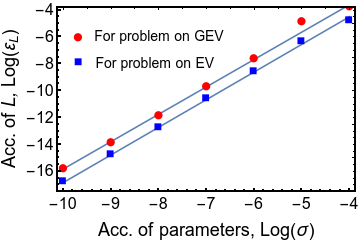}
      \caption{}
    \end{subfigure} 
     \begin{subfigure}{0.4\textwidth}
      \includegraphics[width=0.7\textwidth]{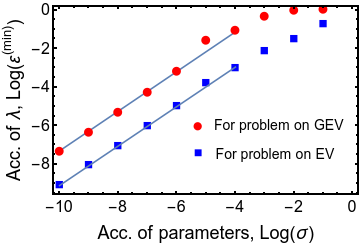}
      \caption{}
    \end{subfigure}  
    \caption{(a) The iteration number $N_r$, (b) the accuracy of the loss function $\log \varepsilon_L$ and  (c) the minimal  reachable accuracy of eigenvalue $\log \varepsilon^{(min)}$ as functions of the accuracy of encoding the optimization parameters $\log \sigma$ for both problems on  GEV and EV. These figures show that all presented quantities  are proportional to $\log \sigma$ for small enough $\sigma$. The red circles and blue squares  correspond, respectively, to the problem on GEV discussed in Sec.\ref{Section:SVD},  and to the problem on EV discussed in Sec.\ref{Section:ev}.  Slopes of both lines in plates (b) and (c)  are very close to each other. Similar to Fig.\ref{Fig:LossF}, the algorithm for constructing  EV converges  faster.}
\label{Fig:LossFPert}
\end{figure}

As mentioned in Sec.\ref{Section:var}, the optimization parameters $\hat \gamma$ can be implemented with the accuracy $\sigma$ which determines the minimal accuracy for the eigenvalues $\varepsilon^{(min)}$. We emphasize that such approximate encoding of optimization parameters (which are parameters of one-qubit rotations) does not beak unitarity of $U_\chi$ and $U_\psi$.
In this case, we break the  optimization algorithm  if  the iterations stop converging  the eigenvalues to their theoretically predicted  values,   i.e.,  the algorithm stops if the condition  $1/4 \sum_{i=1}^4 |\lambda^{(n-1)}_i-\lambda^{(n)}_i| >  10^{-12}$ is destroyed. Here  we use some conventional quantity  $10^{-12}$ characterizing the convergency rate,  $\lambda^{(n)}_i$ denotes the $i$th eigenvalue at the $n$th iteration step. 
 For  the same set of 100 pairs of matrices $\{A,B\}$ (or just of matrices $A$ in the case of the problem on EV), we 
 obtain   the   iteration number $N^{(it)}$, the  logarithm  of the  loss function $L=\varepsilon_L$ and the logarithm of $\varepsilon^{(min)}$ averaged over 100 experiments  as functions of $\log (\sigma)$. Results are shown by the red circles and blue squares for, respectively, the problems on  GEV and EV on Fig.\ref{Fig:LossFPert}.

Fig.\ref{Fig:LossFPert}a demonstrates that the convergence  rate is significantly larger in the case of algorithm for constructing EV.  We emphasize that the slopes of the  lines approximating the sets  of red circles and blue squares in Fig.\ref{Fig:LossFPert}b,c are wery close to each other that confirms Eq.(\ref{ee2}) and Eq.(\ref{ee3}) which hold for both algorithms. 

In this case, $N^{(it)}$, $\log \varepsilon_L$ and $\log \varepsilon^{(min)}$ for the considered family of 100 experiments can be approximated by the linear function of $\log \sigma$ for $\sigma \le  10^{-6}$:
\begin{eqnarray}\label{pertN}
&&
N^{(it)} =   -\varkappa^{(N)} \log  \sigma  -C^{(N)}, \;\; \varkappa^{(N)}=\varkappa^{(N;0)} \pm \Delta \varkappa^{(N)},\;\; C^{(N)}= C^{(N;0)} \pm \Delta C^{(N)},\\\nonumber
{\mbox{for problem on GEV: }} &&\varkappa^{(N;0)}  = 2981.93,\;\;  C^{(N;0)}=13074.7,\;\; 
\Delta \varkappa^{(N)} = 2748.08,\;\; \Delta C^{(N)} =  18606.5,
\\\nonumber
{\mbox{for  problem on EV: }} &&\varkappa^{(N;0)}  = 922.76,\;\;  C^{(N;0)}=3456.64,\;\; 
\Delta \varkappa^{(N)} =  1395.32,\;\; \Delta C^{(N)} =   8746.82,
\\\nonumber \\\label{pertL}
 && \log \varepsilon_L = \varkappa^{(L)} \log \sigma  +C^{(L)} \;\;\Rightarrow \;\; \varepsilon_L = 10^{- C^{(L)}} \sigma^{\varkappa^{(L)}}\\\nonumber
 &&
  \varkappa^{(L)}=\varkappa^{(L;0)}\pm \Delta \varkappa^{(L)}, \;\;  C^{(L)}=C^{(L;0)}\pm \Delta C^{(L)}, \\\nonumber
{\mbox{for problem on  GEV: }} &&  \varkappa^{(L;0)}= 2.034, \;\; C^{(L;0)}= 4.528, \;\; \Delta \varkappa^{(L)} =  0.245,\;\; 
 \Delta C^{(L)}=3.509,
 \\\nonumber
{\mbox{for problem on EV: }} &&  \varkappa^{(L;0)}= 2.053, \;\; C^{(L;0)}=  3.714, \;\; \Delta \varkappa^{(L)} =  0.240,\;\; 
 \Delta C^{(L)}=2.919,
 \\\nonumber \\\label{perte}
 &&
 \log \varepsilon^{(min)}  =  \varkappa^{(\lambda)} \log \sigma  +C^{(\lambda)} \;\; \Rightarrow \;\;  \varepsilon = 10^{- C^{(\lambda)}} \sigma^{\varkappa^{(\lambda)}},\\\nonumber
 && \varkappa^{(\lambda)}=\varkappa^{(\lambda;0)}\pm \Delta \varkappa^{(\lambda)}, \;\;  C^{(\lambda)}=C^{(\lambda;0)}\pm \Delta C^{(\lambda)}, \\\nonumber
{\mbox{for problem on GEV: }} &&  \varkappa^{(\lambda;0)}= 1.028, \;\; C^{(\lambda;0)}= 2.976, \;\; \
 \Delta \varkappa^{(\lambda)}= 0.116,\;\; \Delta C^{(\lambda)}= 2.308,
 \\\nonumber
{\mbox{for problem on EV: }}  &&
  \varkappa^{(\lambda;0)}= 1.024, \;\; C^{(\lambda;0)}=1.131, \;\; \
 \Delta \varkappa^{(\lambda)}= 0.114,\;\; \Delta C^{(\lambda)}= 1.145.
 \end{eqnarray}
  Similar to the $N^{(it)}$ in Sec.\ref{Section:unpert}, the parameters $\varkappa^{(N)}$ and $C^{(N)}$ of linear approximation for  $N^{(it)}$ in Eq.(\ref{pertN})  have large mean-square deviations, so that the approximating lines are completely different for different pairs of matrices $\{A,B\}$ (or for different matrices A in the case of problem on EV). The mean-square deviation of the parameters $\varkappa^{(L)}$ and $\varkappa^{(\lambda)}$ (slopes of the lines) in, respectively, Eqs.(\ref{pertL}) and (\ref{perte}) are relatively small, they take, respectively, $\approx 12\%$ and 
 $\approx 11\%$ of the appropriate mean values for both problems on GEV and EV.  This agrees with formulae (\ref{ee2}) and (\ref{ee3}) for $\varepsilon_L$ and $\varepsilon^{(min)}$, because  $\varkappa^{(L)}\approx 2$ and $\varkappa^{(\lambda)}\approx 1$ in formulae, respectively, (\ref{pertL}) and (\ref{perte}). The mean-square deviations of   $C^{(L)}$ and  $C^{(\lambda)}$ are large, similar to mean-square deviation for $C^{(L)}$ in (\ref{Lappr}), i.e., these parameters significantly depend on the considered  matrices $A$ and $B$.

\section{Conclusions}
\label{Section:conclusions}
We apply  the   method of matrix encoding into the quantum  state to the  quantum algorithm of calculating the loss function in frames of the variational algorithm for  constructing of  both GEV and  EV.  The matrices $A$ and $B$ under consideration are encoded into the superposition state of the subsystems $R$, $C$ and $L$, at that the state of the one-qubit subsystem $L$ labels the parts of the superposition state associated with the matrices $A$ and $B$. This algorithm is applicable to arbitrary pair of complex  square matrices $A$ and $B$. However, in this paper,  we consider the non-degenerate matrices and use the lower boundary $0.1$ for  the numerical simulation of GEV and EV.  We remove the garbage acquired during calculations via measurement of the state of  the ancilla  $B$  with the probability of desired output (success probability) $\sim N^{-4}$.  The depth of the circuit is estimated by either $O(M\log N)$ or $O(N^2\log N)$ depending on $M$, while the space is $O(\log  N)$. We emphasize that the  estimation  $O(N^2\log N)$ is related to selecting the vanishing elements from the matrices $T$ and $S $ in  expression (\ref{obj}) for the  loss function. 

After removing the garbage, the loss function can be obtained probabilistically performing the set of  $\sim1/\varepsilon_L$ runs so that the loss function becomes expressed in terms of the probability amplitudes of the superposition state which originally contained  the elements of the matrices $A$ and $B$, see Eqs.(\ref{Phi4}) - (\ref{Lp}). The number of runs needed to calculate the loss function together with all necessary  derivatives with respect to the  optimization parameters is $\sim M\log N /\varepsilon _L$, where  $M\log N$ is the estimated number of optimization parameters and the accuracy $\varepsilon_L$ of $L$ depends on the desired accuracy $\varepsilon$  of GEV (or EV). We show that $\varepsilon_L  \sim \varepsilon^2$.  The accuracy $\sigma$ of encoding the optimization  parameters   imposes the lower boundary $\varepsilon^{(min)}$ on the accuracy of the calculated eigenvalues and $\varepsilon^{(min)}\sim \sqrt{\varepsilon_L}\sim \sqrt{\sigma}$. The above theoretically obtained relations among accuracy-parameters 
$\varepsilon$, $\varepsilon_L$ and $\sigma$ are confirmed by numerical simulations with 100 pairs of  square random complex $4\times 4$ matrices $A$ and $B$ for the problem on GEV and with 100 matrices $A$ for the problem on EV. The advantage of matrix-encoding method is that it is applicable to any matrices (subjected to normalization (\ref{A})) and does not require representing the   input matrices as linear combinations of unitary matrices or embedding them   into the Hermitian matrix  with subsequent exponentiating to obtain a unitary operator. 

One also has to remember the success probability  on the step of garbage removing which is $O(N^{-4})$ and  creates certain problem. But our expectation is that such problem having quantum nature can be overcome via quantum methods. A possible variant might be the  so-called controlled measurement \cite{FZQWarxive2025}, which would allow to resolve the problem of small success probability with minimal expenses. However, this method still requires justification.

{\bf Acknowledgments.} The work of J.Wu was  supported by the National Natural Science Foundation of China ( (Grants No.
12271474, No. 12031004 and No. 61877054)). The work of A. I. Z. was  funded by a state
task of Russian Fundamental Investigations (State Registration No. 124013000760-0).

\end{document}